\documentclass[letterpaper,journal]{IEEEtran}
\usepackage{amsmath,amsfonts}
\usepackage{algorithmic}
\usepackage{algorithm}
\usepackage{array}
\usepackage{textcomp}
\usepackage{stfloats}
\usepackage{url}
\usepackage{verbatim}
\usepackage{graphicx}
\usepackage{tabularx}
\usepackage{subcaption}
\usepackage{cite}
\usepackage{bm}
\usepackage{threeparttable}
\usepackage{amssymb}
\usepackage{booktabs}
\usepackage{tabularray}
\usepackage{caption}
\usepackage{textcomp}
\usepackage[T1]{fontenc}

\def\BibTeX{{\rm B\kern-.05em{\sc i\kern-.025em b}\kern-.08em
    T\kern-.1667em\lower.7ex\hbox{E}\kern-.125emX}}

\usepackage{balance}

\begin{document}
\title{Shatter Throughput Ceilings: Leveraging Reflection Surfaces to Enhance Transmissions for Vehicular Fast Data Exchange}

\author{Qianyao Ren,
Qingxiao Huang,
Yiqin Deng, ~\IEEEmembership{Member,~IEEE},
Xianhao Chen, ~\IEEEmembership{Member,~IEEE},
Phone Lin, ~\IEEEmembership{Fellow,~IEEE},
and Yuguang Fang, ~\IEEEmembership{Fellow,~IEEE}
\IEEEcompsocitemizethanks{
    \IEEEcompsocthanksitem Qianyao Ren, Qingxiao Huang, and Yuguang Fang are with the Hong Kong JC STEM Lab of Smart City and the Department of Computer Science, City University of Hong Kong, Kowloon, Hong Kong \protect(E-mail: qianyaren2-c@my.cityu.edu.hk, qx.huang@my.cityu.edu.hk, my.fang@cityu.edu.hk).
    \IEEEcompsocthanksitem Yiqn Deng is with the School of Data Science, Lingnan University, Tuen Men, Hong Kong, Hong Kong \protect(E-mail: yiqindeng@ln.edu.hk).
    \IEEEcompsocthanksitem Xianhao Chen is with the Department of Electrical and Electronic Engineering, University of Hong Kong, Pok Fu Lam, Hong Kong \protect(E-mail: xchen@eee.hku.hk).
    \IEEEcompsocthanksitem Phone Lin is with the Department of Computer Science and Information Engineering, National Taiwan University, Taipei, Taiwan \protect(E-Mail: plin@csie.ntu.edu.tw).
    }
\thanks{The research work described in this paper was conducted in the JC STEM Lab of Smart City funded by The Hong Kong Jockey Club Charities Trust under Contract 2023-0108. The work described in this paper was also partially supported by a grant from the Research Grants Council of the Hong Kong Special Administrative Region, China (Project No. CityU 11216324). }
}

\maketitle

\begin{abstract}
Rapid emergence of smart mobility necessitates high-volume bursty data transmission over a single link between a target vehicle and its designated edge computing-enabled Base Station (BS) or Roadside Unit (RSU), which must be completed within a short time period when the vehicle traverses the coverage area. However, in bandwidth-limited scenarios, conventional communication systems face a fundamental throughput ceiling at each single vehicle. This limitation persists even when all time–frequency resources are allocated to a single vehicle, as the underlying channel lacks sufficient spatial diversity to support higher data rates. To break this throughput ceiling, in this paper, we propose a novel reflection-enhanced transmission framework by strategically employing dedicated specular reflecting surfaces along roadways to proactively augment the transmission environments. This setup concentrates dispersed signals from multiple directions toward a target vehicle, analogous to the light-focusing effect of a concave magnifying lens, thereby enhancing the spatial diversity and achievable rank of an individual channel. This allows a BS to allocate more transmission layers to one single user, consequently significantly raising the throughput ceiling for individual vehicles. Moreover, we also introduce dynamic virtualization methods for reflecting panel patch groups, compatible with existing communication systems, to flexibly manage interference with other coexisting users. Furthermore, collaborative rotation among multiple reflecting panels is introduced to enhance signal concentration. Finally, the schematic effectiveness is rigorously validated through 3GPP-compliant system-level simulations, demonstrating significant throughput boosts.
\end{abstract}


\section{Introduction}
\label{intr}
\IEEEPARstart{T}{he} proliferation of intelligent vehicles has brought about new functionalities that demand higher data transmission rates and shorter latency for vehicular communications systems \cite{Vaas}. These emerging vehicular applications demand completion of substantial data exchange when a vehicle traverse the coverage area of a specific base station (BS), challenging conventional communication systems with link sensitivity and traffic burstiness. Link sensitivity arises as this data exchange primarily occurs on one specific link between a target vehicle (TV) and a designated edge computing device, like a roadside unit (RSU) or a compute-capable BS, rather than a cloud server to minimize the end-to-end latency. Traffic burstiness characterizes the event-driven and time-deficient transmissions where sporadic data volumes can significantly exceed baseline traffic levels and data must be fully transmitted during the traversal period as it rapidly become obsolete once vehicles exit designated service areas.


As a critical function for Autonomous Vehicles (AVs), High-Definition Maps (HD Maps) necessitate swift transmission of bursty data on one specific link to support high-resolution, real-time map updates \cite{HD, HD_com}. For instance, during road emergencies, to avoid the severe data overwhelming and excessive resource consumptions caused by simultaneous transmission \cite{HD_update}, a targeted transmission from a source vehicle at a precise location to a designated BS/RSU is prefered \cite{HD_one}. Link sensitivity and traffic burstiness are equally critical for transmitting high-resolution video data and sensing data in AV and Advanced Driver-Assistance Systems (ADAS) \cite{AV, LiDar, LiDarPDP}. Similarly, for infotainment applications such as Augmented Reality Head-Up Displays (AR-HUDs), high-precision positioning data must be exchanged between reference vehicles and a specific RSU \cite{AR-Hud}. Moreover, the transfer of training datasets and model parameters for artificial intelligence (AI) models employed in driving systems \cite{AV} or communication modules \cite{Twin} can induce bursty traffic within certain time intervals. Additionally, in Store-Carry-Forward data transportation systems \cite{Vaas}, the fast data exchange between a vehicle and the approaching BS further underscores the importance of these two challenges.




The bursty traffic requirement on a single link exceeds the design scope of conventional 5G cellular communication systems \cite{5G, 5G_mobile, HD}. Current 5G systems prioritize maximizing aggregate cell throughput over guaranteeing per-link throughput, making them inadequate for single-link high-throughput scenarios. Although vehicles can switch BSs to maintain connectivity in unfavorable wireless conditions or when moving beyond coverage, it fundamentally conflicts with the link sensitivity and traffic burstiness. Consequently, how to enable a high-volume bursty data transmission over a single link when vehicles pass through BS/RSU coverage zones is the core research problem addressed in this paper.

The data transmission rate is determined by three key factors: modulation and coding scheme (MCS), allocated bandwidth, and the number of MIMO layers. High vehicular mobility induces Doppler frequency shift, which complicates signal detection and degrades the Signal-to-Interference-plus-Noise Ratio (SINR). This shift can be exacerbated at high carrier frequencies, making sub-6 GHz bands preferable for mobile scenarios. Although communication systems like 5G support higher carrier frequencies, current deployments predominantly utilize sub-6 GHz bands \cite{BW_CMCC, BW_Unicom, BW_5G}. This low carrier frequency limits the available bandwidth to about one-quarter of the high-frequency scenarios \cite{3GPP101}, creating a bandwidth-constrained scenario.

Massive MIMO represents a pivotal technology for enhancing transmission rates, enabling simultaneous transmission of multiple independent data streams over identical time-frequency resources \cite{5G_MIMO,3GPP214}. While a BS supports a high maximum number of MIMO layers, the layers actually allocated to an individual user is severely restricted, due to limited spatial diversity or a small user-side antenna array \cite{MIMO_corr}. Traditional communication systems address this limitation through multi-user scheduling, where BSs dynamically assign spatial resources across numerous users to exploit the full multiplexing capability. While this approach maximizes aggregate cell throughput, individual users remain constrained by limited spatial layers, especially vehicular users with high mobility.

Given the constraints of available bandwidth and MIMO layers, a hard throughput ceiling exists for individual TVs. Even when all time-frequency resources are exclusively allocated to a single TV and the receiving SINR is sufficiently high for the maximal MCS, the individual throughput remains inadequate for emerging vehicular applications. This bottleneck is further aggravated by traffic burstiness, which precludes prolonged transmission durations, and by link sensitivity, which limits the spatial multiplexing gains typically achievable in conventional multi-user systems.

\begin{figure}
    \centering
    \includegraphics[width=3.4in]{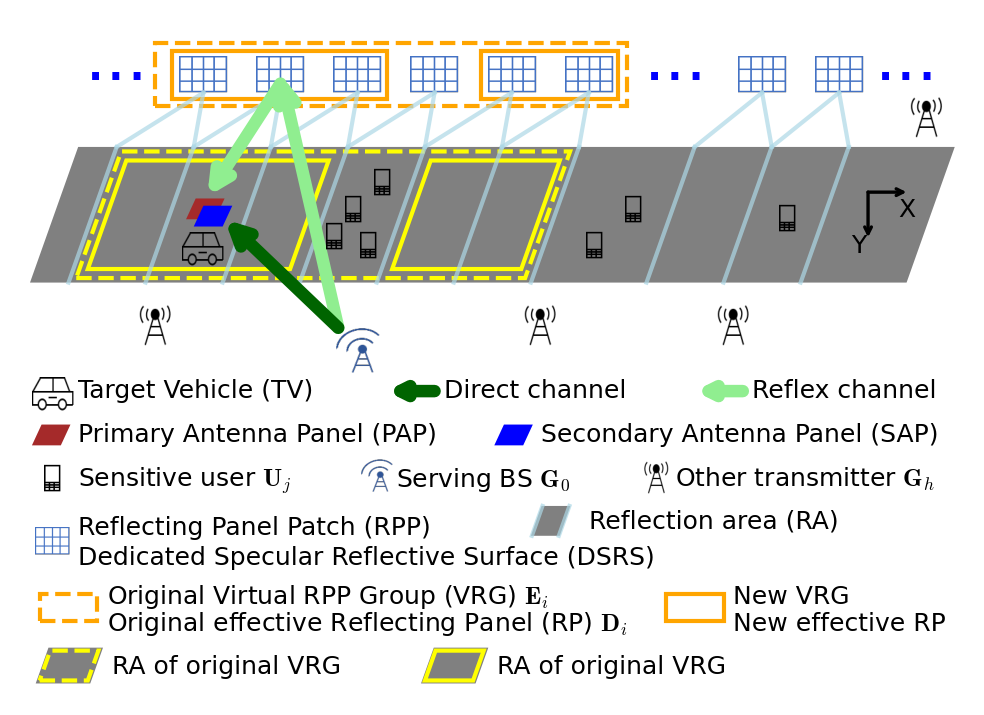}
    \caption{The deployment of reflecting panel patches (RPPs) in the reflection enhanced transmission framework (RETF)}
    \label{Demo}
\end{figure}

To shatter this throughput ceiling, we propose to augment spatial channel diversity for individual vehicles through strategically inducing reflective radio paths, thereby unlocking additional MIMO layers. This paper introduces a method employing a series of Dedicated Specular Reflective Surfaces (DSRS) strategically positioned roadside opposite BSs/RSUs. Analogous to a concave mirror reflecting and focusing daylight onto a single point to burn a paper, these surfaces can create reflective radio paths with multiple directions towards a given TV to boost the received power, as illustrated in Fig. \ref{Demo}. These additional reflective paths permit the TV to activate antenna panels not directly oriented toward the BS, thereby expanding the channel dimensionality and increasing its maximum achievable rank. The spatial integration of these diverse paths also significantly improves channel diversity, which elevates the probability of the practical rank approaching the theoretical maximum. These two improvements enable the BS to schedule more MIMO layers for the TV, ultimately raising its throughput ceiling. Within this compatible framework, the BS can substantially boost the data transmission rate for high-priority TVs with traffic burstiness and link sensitivity requirements. When there are no such TVs, the system seamlessly reverts to conventional multi-user scheduling mechanisms, ensuring fair and efficient resource allocation among normal users.


In this paper, we start with introducing DSRS, a low-cost alternative to the traditional Intelligent Reflecting Surface (IRS). With the utilization of the Reflecting Panel (RP), a practical implementation of DSRS, we establish the system model. Next, we introduce the State-Switching Mechanism (SSM) and Reflecting Panel Patches (RPPs) managed by a Dynamic Group Virtualization (RPP-DGV) system, designed to maximize the transmission rate of TVs while minimizing interference with coexisting users. Leveraging a Greedy Stepwise Search Algorithm (GSSA), we develop a fast virtualization scheme and present the Reflection-Enhanced Transmission Framework (RETF). To further boost throughput, we propose the Rotational Cooperative Team (RCT) and Alternating Neighbor Selection (ANS) mechanism to make multiple reflex signals focus on a specific region. Finally, we develop a System-Level Simulation (SLS) platform, based on 3GPP TR 38.901 \cite{3GPP901}, to quantitatively evaluate RETF.

\section{Dedicated specular reflective surfaces}
IRS have garnered significant attention as a promising technology for 6G communication networks. IRS can enhance wireless transmissions by intelligently configuring the phase shift of each reflecting element, thereby improving system capacity and coverage \cite{RIS_principle, RIS_survey, Reflector_tuned}. The utilization of IRS can provide alternative paths to mitigate blockage and enhance coverage \cite{RIS_blockage}, and assist millimeter-wave (mmWave) MIMO systems by connecting more BSs and users to increase overall capacity \cite{RIS_assist}. Generally speaking, if intelligently utilized, conventional IRS can be deployed along the road to augment the wireless environment and enhance spatial diversity. 


However, directly applying the intelligent phase control to emerging vehicular applications is analogous to using a sledgehammer to crack a nut. These applications, as outlined in Section \ref{intr}, are predominantly utilized in low-frequency, high-mobility, and single-link scenarios where the sophisticated intelligence of IRS becomes excessive and redundant.

IRS effectively mitigates the severe path loss in high-frequency communications by performing dynamic optimization over numerous element coefficients \cite{LIS_assist, RIS_beam}. Whereas, in low-frequency scenario characterized by rich multi-path propagation, even a simple, non-intelligent reflection as illustrated in Fig. \ref{Demo} can substantially enhance the spatial multiplexing capacity of the MIMO system in a targeted area. The potential of precise phase reconfiguration remains underutilized, which is further constrained by the impractical physical aperture size required for low-frequency ranges.

The intelligent phase reconfiguration of an IRS fundamentally depends on accurate Channel State Information (CSI) \cite{LIS_assist, RIS_coeff}. In high-mobility scenarios, however, CSI becomes quickly outdated due to rapidly changing vehicle locations and channel conditions, necessitating frequent channel detections and CSI. In the multi-path environments of sub-6 GHz bands, CSI estimation and CSI feedback consume substantial radio resources \cite{CSI_type2, CSI_AI} and introduce interference to other users, thereby degrading the overall system throughput.

Similar to conventional 5G systems, a predominant focus of these IRS-assisted MIMO systems is on maximizing aggregate cell throughput, typically by employing multiple IRS and BSs to serve many spatially separated users \cite{LIS_assist, RIS_assist}. Due to the geographical dispersion, the channels to these client users are largely independent, effectively forming a high-rank composite channel. In single-link scenarios, however, the multi-antenna TV cannot be equated to multiple independent single-antenna users since antenna correlations often lead to a low-rank channel \cite{MIMO_corr}. Thus, effectively improving MIMO gain for this specific single-link communication between the TV and its serving BS remains an open research.

Therefore, for the emerging vehicular applications, only partial capabilities of an IRS is effectively utilized, rendering it to \textquotedbl non-intelligent\textquotedbl . Conversely, the significant deployment costs, high radio resource consumption, and substantial computational complexity incurred by its intelligent functionalities present a major barrier to practical adoption. To address this problem, we need a new type of \textquotedbl non-intelligent\textquotedbl \space IRS, which is inherently low-cost and resource-efficient, as a practical alternative to the conventional IRS.

Recently, fully passive reflectors, essentially metallic plates, have been introduced into communication systems to enhance coverage \cite{Passive_coverage} and improve MIMO channel rank \cite{Passive_Deployment}. In \cite{Reflector_tuned}, we proposed tunable reflectors comprising highly reflective metallic plates that are mechanically adjusted to create alternative Line-of-Sight (LoS) paths in Wireless Local Area Network (WLAN) systems. Building on the concept of movable antennas \cite{Movable_Antenna}, the flexible reflector presented in \cite{Flexible_Reflector} can enhance coverage by adjusting its position and orientation to match the specular direction. Meanwhile, analysis and experiments in \cite{Metal_Reflecting} reveal that fully passive reflectors can effectively enhance a broader region centered on the specular direction.

\begin{figure}
    \centering
    \includegraphics[width=3.4in]{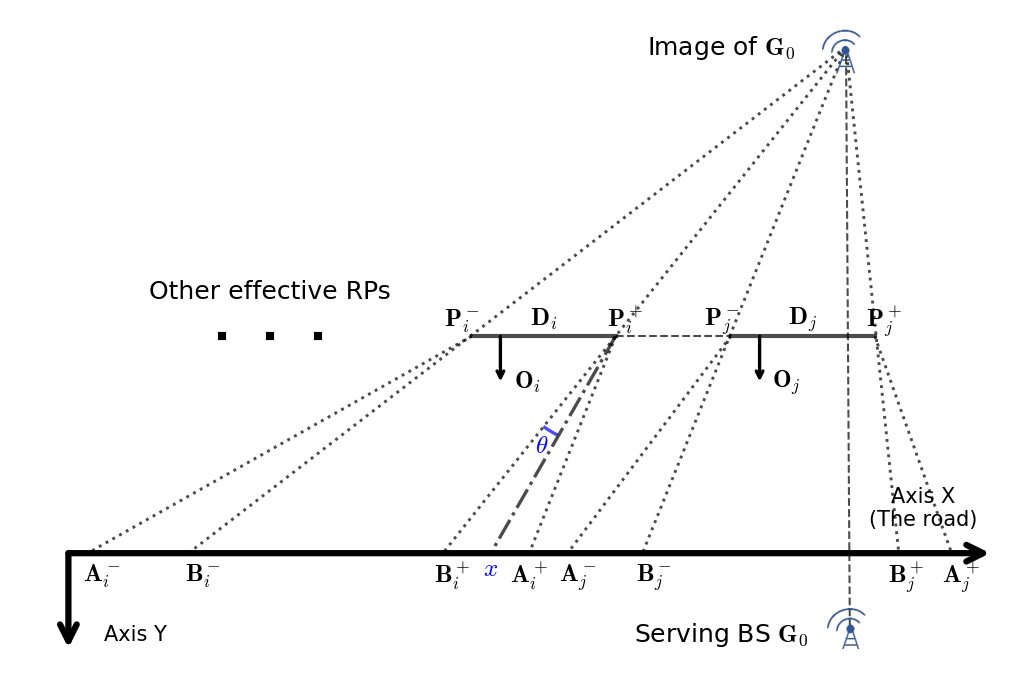}
    \caption{The reflection areas (RAs) of the reflecting panels (RPs)}
    \label{RA}
\end{figure}

Although these works focused on mmWave frequencies, the idea of mechanically adjusted reflective plates remains applicable to lower-frequency bands. In this paper, we propose DSRS as a cost-effective IRS paradigm operating purely via specular reflection without phase manipulation. DSRS can be realized as arbitrary physical objects with geometrically predictable reflection areas (RAs). Given that multi-path wireless environments are more intricate in lower-frequency mobile scenarios, the specular reflection region, illustrated as region $\left[B^-_i, B^+_i\right]$ in Fig. \ref{RA}, can achieve near-maximal received power. Each DSRS is dedicated to a fixed RA, indicated by a distinct gray parallelogram in Fig. \ref{Demo}, thereby establishing a one-to-one mapping between a geographical region and its corresponding DSRS. Within its dedicated RA, DSRS emphasizes on augmenting transmission environments by increasing diverse paths rather than adapting them with phase reconfiguration. 

The RA of each DSRS is determined solely by its relative position and orientation with respect to the transmitter, requiring no knowledge of the wireless environment. This fundamentally eliminates the need for precise CSI acquisition. A BS only needs the approximate location of a TV to configure the corresponding DSRS, which is available since TVs follow predetermined routes along the road. Additionally, this one-to-one association renders the DSRSs mutually independent, which enhances deployment flexibility. Once the minimum reflection size is satisfied, the physical size and permissible orientation of each DSRS, along with the inter-DSRS spacing, are all adjustable according to the conditions of the target deployment environment. Although the mechanical control of DSRS introduces minor latency, a BS can configure DSRS orientations in advance  based on the predicted TV trajectories, ensuring readiness when a TV enters the specific area.

In this paper, reflecting Panels (RPs) are proposed as the practical realization of DSRSs. The RPs are rotatable plastic panel coated with multilayered conductive materials for high electromagnetic reflectivity. Each RP is integrated with a miniature stepping motor, enabling precise rotation as the mechanically driven MAs with rotatable shafts in \cite{motor}. In practice, RPs can be installed on advertising bulletin boards, trees, street lamps, and building walls beside the road or mountains along highways. Their designs can be customized in shape and size to blend into diverse environments without damaging their aesthetics.



\section{System Model}


\subsection{Fundamental Topological Structure}
\label{Topology}
In this paper, we analyze a straight road in which the path of a TV corresponds to the X-axis, with the positive direction aligned with the movement of the TV, as illustrated in Fig. \ref{Demo}. The serving area of a BS along the road is defined by the interval $[0,L]$. There are a total of $N_G+1$ transmitters represented as
\begin{equation}
    \mathcal{G}=\{\bm{G_h}=\left(\bm{Q}_h,\bm{A}_h, \overline{s}_h\right),\ h=0,1,2\dots N_G\}
\end{equation}
where $\bm{Q}_h$ indicates the 3-D location of the $h$-th transmitter $G_h$, $\bm{A}_h$ denotes the orientation of its antenna panels, and $\overline{s}_h$ represents the transmit power. The first transmitter $G_0$ with $(\bm{Q}_0,\bm{A}_0, \overline{s}_0)$ represents the serving BS for the TV, whereas the others function as interference transmitters. The TV enters this interval at time $t=0$ and exits at time $t=T$ with its position recorded as $\bm{x_t}=(x_t, 0, 0)$, where $x_t$ is the location of the TV on X-axis at time $t$.

We assume that $N_D$ effective Reflecting Panels (RPs), illustrated as orange rectangles in Fig. \ref{Demo}, are deployed along the roadside opposite $\bm{G_0}$. Each effective RP is a composite structure comprising a series of successive reflecting panel patches, as detailed in Section \ref{RPP_Groups}. The set of these effective RPs is denoted by $\mathcal{D}$, which is defined as
\begin{equation}
    \mathcal{D}=\{\bm{D}_i=\left(\bm{P}_i^-,\bm{P}_i^+,\bm{O}_i\right),\ i=1,2...N\}
\end{equation}
where $\bm{P}_i^-$ and $\bm{P}_i^+$ indicate the starting and ending 3-D location of the $i$-th RP $\bm{D}_i$, respectively. $\bm{O}_i=(\cos{\alpha_i},\sin{\alpha_i},\cot{\gamma_i})$ represents the orientation of $D_i$ in 3-D coordinates with azimuth angle $\alpha_i$ and zenith angle $\gamma_i$. Taking into account that vehicles solely travel along the road and do not move across the road, we concentrate our attention on the horizontal RA. We assume that each RP sufficiently covers the entire road in the vertical direction with a suitable $\gamma_i$.


We define sensitive users (SUs) as those whose transmission performance is degraded by reflected signals. These SUs are connected to different BSs or operate within alternative communication systems, which precludes the serving BS from scheduling them idle to mitigate reflected interference when the TV moves past them. We assume that each mobile SU exhibits uniform motion and utilize $\kappa$ to denote the ratio of mobile SUs. We postulate that all SUs experience full buffer traffic and that their locations are randomly generated along the road, represented as
\begin{equation}
    \mathcal{U}=\{\bm{U}_j=(x_j,y_j,h_j,v_j),\ j=1,2\dots N_U\}
\end{equation}
where $x_j$ indicates the location of the $j$-th SU $\bm{U}_j$ along the X-axis and $\bm{x}_j=(x_j,y_j,0)$ denotes its 3-D coordinates. $y_j$ denotes a random distance to the moving route of a TV, which is no more than the width of the road. For mobile SUs, these locations are defined as the encounter point with the TV. $h_j\in\{1,2\dots N_G\}$ represents the index of the transmitter designated as the serving transmitter for $\bm{U}_j$. $v_j$ indicates the speed of $\bm{U}_j$ and for a stationary SU, $v_j=0$. $N_U$ denotes the number of SUs. 

\subsection{LOS Angle and Antenna Pattern}
\label{LOS_Angle}
For direct channels, illustrated as the arrow with deep color in Fig. \ref{Demo}, LOS angles of arrivals and departures are dictated by the positions of transmitters and TVs. In contrast, for reflex channels, illustrated as arrows with light color in Fig. \ref{Demo}, the LOS angles of arrivals are determined by the locations of TVs and RPs, while the locations of RPs and the serving BS influence the LOS angles of departures.

To simulate the plentiful antenna panels oriented at various directions on intelligent vehicles, we consider a typical antenna configuration comprising two opposing panels positioned 180 degrees apart, each equipped with an equal number of antennas. These two antenna panels, the Primary Antenna Panel (PAP) and the Secondary Antenna Panel (SAP), are oriented toward the BS side and the RP side, denoted as $\bm{B}^{(p)}$ and $\bm{B}^{(s)}$, respectively.

For these two antenna panels at TVs and those at transmitters, we employ a directional antenna pattern referenced in \cite{3GPP901}. The power gain can be derived from Table 7.3-1 in \cite{3GPP901}, expressed as $g(\bm{P}_A,\bm{P}_B,\bm{O}_A)$, where $\bm{P}_A$ and $\bm{O}_A$ denote the location and the orientation of the antenna panel, respectively and $\bm{P}_B$ denotes an arbitrary point on the radio link to indicate the observed direction. The discrepancy between the orientation of the antenna panel and the direction of the incoming waves can significantly reduce the power gain. Consequently, with respect to the serving BS, these two antenna panels can be assumed to be independent. The direct channel is associated with PAP, whereas the reflex channel is tied to SAP. Due to the random distributions of the SUs, we assume that their antennas are omnidirectional to meet sensitivity requirements, with their power gain set to $1$.


\subsection{Power Loss Model}
\label{loss}
Due to the long distance between the transmitters, RPs, and TVs, we utilize a far-field model that incorporates plane wave characteristics for both incident and reflex paths \cite{3GPP901}. Reflection-induced power loss primarily depends on the propagation distances and the Radio Cross Section (RCS) \cite{Flexible_Reflector, Metal_Reflecting}. Accordingly, we follow the statistical loss model in TR 38.901 \cite{3GPP901} for the distance-related loss and introduce an area-based loss for the RCS-related loss.

\subsubsection{Path loss (PL)}
As provided in TR 38.901 \cite{3GPP901}, the PL depends on the carrier frequency and the distance between the starting point ($\bm{P}_A$) and ending point ($\bm{P}_B$) of a link, denoted as $d(\bm{P}_A, \bm{P}_B)$. The corresponding power ratio $\omega_{PL}(\bm{P}_A, \bm{P}_B)$ can be obtained by $PL=-10\log_{10}(\omega_{PL})$. For reflex channel, $\omega_{PL}^{(ref)}(\bm{P}_A, \bm{P}_B)$ is derived based on the sum of the two-staged distance formed by $d(\bm{P}_A, \bm{P}_*)$ and $d(\bm{P}_*, \bm{P}_B)$, where $\bm{P}_*$ represents the 3-D location of the geometrical reflecting point on the corresponding RP.

\subsubsection{Reflection loss (RL)}
The power ratio $\omega_{RL}$ models the combined effect of maximum RCS gain and the absorption loss for an RP. It is constant for all RPPs of a given size (Section \ref{RPP_Groups}). The RL is defined as $RL=-10\log_{10}(\omega_{RL})$.


\subsubsection{Bias Loss (BL)}
\label{bias_loss}
Following the experiments in \cite{Metal_Reflecting}, we model the specular reflection region, where the angle of incidence equals the angle of reflection, with a constant RCS and a unit power ratio. Users outside this region may experience a significant power reduction, which we define as the BL. The corresponding power ratio is denoted by $\omega_{BL} \leq 1$. 


Fig. \ref{RA} presents the effective RP $\bm{D}_i$ and its RA within the XOY plane. The two endpoints of $\bm{D}_i$ is denoted as $\bm{P}_i^-=\left(P_{i,x}^-,P_{i,y}\right)$ and $\bm{P}_i^+=\left (P_{i,x}^+,P_{i,y}\right)$. Azimuth angle $\alpha_i$ indicates the orientation of $\bm{D}_i$ on XoY plane. If $\bm{D}_i$ is aligned with the road without rotation, $\alpha_i=90^\circ$.

The corresponding mirror reflecting region can be established by introducing the mirror image of BS on $\bm{D}_i$, identified as $[B_i^-, B_i^+]$ on the X-axis. This specific region is termed as the Direct Reflection Area (DRA), which ensures vehicles situated within this designated area to experience no additional BL. The DRA can be calculated based on the geometrical relationship with $B^-_{i,x}$ expressed as
\begin{equation}
    B^-_{i,x}=\frac{Q_{0,y}-2P_{i,y}}{Q_{0,y}-P_{i,y}}P_{i,x}^-+\frac{Q_{0,x}P_{i,y}}{Q_{0,y}-P_{i,y}}
\end{equation}
where $\bm{\overline{Q}}_0=(Q_{0,x},Q_{0,y})$ is the location of the BS on the XoY plane. Similarly, $B^+_{i,x}$ can be determined by $P_{i,x}^+$.
 
Due to the angular spread of the channel from transmitters to RPs, the actual incident angle can extend beyond the geometrical incident angle. This gives rise to Indirect Reflection Areas (IDRA), denoted as $\left[A_i^-, B_i^-\right]$ and $\left[B_i^+, A_i^+\right]$, where vehicles experience an additional location-dependent power loss but can still receive reflected signals. 

As per Eq. (7.5-11) in \cite{3GPP901}, the angle $\theta$ in Fig. \ref{RA} follows a zero-mean Gaussian distribution. Given that the IDRA length is much smaller than the distance from the effective RP to the road, it is reasonable to approximate $\left|x-B_i^+\right|\propto \theta$ for a vehicle at position $x$ within $\left[B_i^+, A_i^+\right]$. The corresponding power ratio factor $\omega_{BL}(i,x)$ is modeled as
\begin{equation}
\label{IDRA}
    \omega_{BL}(i,x) = e^{-\chi\left|x-B_i^+\right|^2}, \quad x\in \left[B_i^+, A_i^+\right]
\end{equation}
where $\chi$ denotes a constant decaying factor, defined by the angle spread, following the random variation in Section 7.5 of \cite{3GPP901}. A symmetric expression applies to the opposite region $\left[A_i^-, B_i^-\right]$. We then define the full region $\left[A_i^-, A_i^+\right]$ as the RA of $\bm{D}_i$, which comprises both the DRA and the IDRA.

It is assumed that a vehicle situated outside this region is unable to receive effective reflection waves from $\bm{D}_i$ because the power gain falls below a specified threshold, $\omega_0\in[0,1]$. A unit power ratio without BL is assigned to the DRA, while the power ratio across the IDRA is governed by \eqref{IDRA}. Thus, the approximate length of the RA, which consists of a single DRA and two IDRAs, is expressed as
\begin{equation}
    \xi_i=\frac{Q_{0,y}-2P_{i,y}}{Q_{0,y}-P_{i,y}}\eta_i+2\sqrt{\frac{-\log\omega_0}{\chi}}
\end{equation}
where the first term is the length of DRA and the second part is the length of the two IDRAs. $\eta_i=P_{i,x}^+-P_{i,x}^-$ is the length of $\bm{D}_i$.

For the gap between two adjacent RPs, denoted as region $\left[A_i^+, A_{i+1}^-\right]$, as illustrated in Fig. \ref{RA}, the power ratio is zero because none of the RPs are available to provide reflection in this interval. If two RPs are positioned closer together, with $A_{i+1}^-$ situated to the left of $ A_i^+ $, the total power gain for the region $\left[A_{i+1}^-, A_i^+\right]$ is the sum of two power ratios from each RP. In general, for a specific location $x$, the BL power ratio can be expressed as:
\begin{equation}
\label{eq:two_rp}
    \omega_{BL}(x)=\min\left(\sum_{i=1}^{N_D}\omega_{BL}(i,x), 1\right)
\end{equation}
When the RPs are deployed, we can accurately calculate the BL for each location $x$ as $BL=-10\log_{10}(\omega_{BL}(x))$.

\subsection{Fast Fading Channel Model}
\label{Model}
In this paper, we employ the fast fading model outlined in TR 38.901 \cite{3GPP901}. This model composes of a Large Scale Model (LSM) for overall received power and a Small Scale Model (SSM) for per-ray distributions of power, delay and angle. All parameters are generated randomly based on their predefined distributions.

We independently generate Channel Impulse Responses (CIRs) in the time domain for the direct channel and reflex channel. Following the procedure in Section 7 of \cite{3GPP901}, CIRs of the direct channel is expressed as
\begin{equation}
    h^{(dir)}(t)=\left\{h^{(dir)}(t,\tau_u)\in\mathbb{C}^{N_r\times N_t}|u=1,2\dots N_{ray}^{dir}\right\}
\end{equation}
where $N_r$ is the number of antennas on each antenna panel at the vehicle side, and $N_t$ is the number of antennas at the BS side. $N_{ray}^{dir}$ is the number of rays of the direct channel. For the reflected channel $h^{(ref)}(t)$, the total PL in LSM phase incorporates RL, BL and the two-stage PL discussed in Section \ref{loss}, while ray angles in SSM phase are generated based on the adjusted LOS angles discussed in Section \ref{LOS_Angle}.

To combine these two channels, we transform $h^{(dir)}(t)$ and $h^{(ref)}(t)$ into frequency domain with the Fast Fourier Transform (FFT), obtaining the corresponding frequency domin channels, $H^{(dir)}_v\in\mathbb{C}^{N_r\times N_t}$ and $H^{(ref)}_v\in\mathbb{C}^{N_r\times N_t}$, where $v=1,2\dots N_B$ and $N_B$ denotes the number of subcarriers. The joint channel in frequency domain can be expressed as
\begin{equation}
    H_v=\begin{bmatrix}H_v^{(dir)} \\ \sqrt{\omega_*} H_v^{(ref)}\end{bmatrix},v=1,2\dots N_B
\end{equation}
where $\omega_*$ is a power factor to adjust the amplitude of the reflex channel matrix to compensate for the disparity in power because the receiving power of the reflex channel is inferior to that of the direct channel. For a specific location $x_t$, the power factor $\omega_*$ can be expressed as
\begin{equation}
    \omega_*=\frac{g^{(dir)}_{a}(x_t) \times\omega_{PL}(\bm{x}_t,\bm{Q}_0)}{g^{(ref)}_{a}(x_t) \times\omega_{PL}^{(ref)}\left(\bm{x}_t,\bm{Q}_0\right) \times \omega_{RL} \times \omega_{BL}(x_t)}
\end{equation}
where $g^{(dir)}_{a}(x_t)=g\left(\bm{Q}_0,\bm{x}_t,\bm{A}_0\right) \times g\left(\bm{x}_t,\bm{Q}_0,\bm{B}^{(p)}\right)$ and $g^{(ref)}_{a}(x_t)=g(\bm{Q}_0,\bm{P}_*(\bm{x}_t),\bm{A}_0) \times g\left(\bm{x}_t,\bm{P}_*(\bm{x}_t),\bm{B}^{(s)}\right)$ are the antenna pattern gain at the transmitter and the receiver for the direct channel and the reflex channel, respectively. $\bm{P}_*(\bm{x}_t)$ indicates the corresponding geometrical reflecting point.

As for the channels between the TV and other interference transmitters ($\bm{G}_h,h=1,2\dots N_G$), each transmitter corresponds to either PAP or SAP according to which side of the road it is located on. We directly generate these channels following the procedure of Section 7 in \cite{3GPP901}, denoted as $H_v^{(h)}$ where $h$ is the index of the transmitters and $v$ is the index of the subcarrier. Similarly, the channels corresponding to the SUs ($\bm{U}_j,j=1,2\dots N_U$) can be represented as $H_v^{(h,j)}$, where $j$ is the index of the SU.

\begin{figure*}
    \centering
    \includegraphics[width=6.5in]{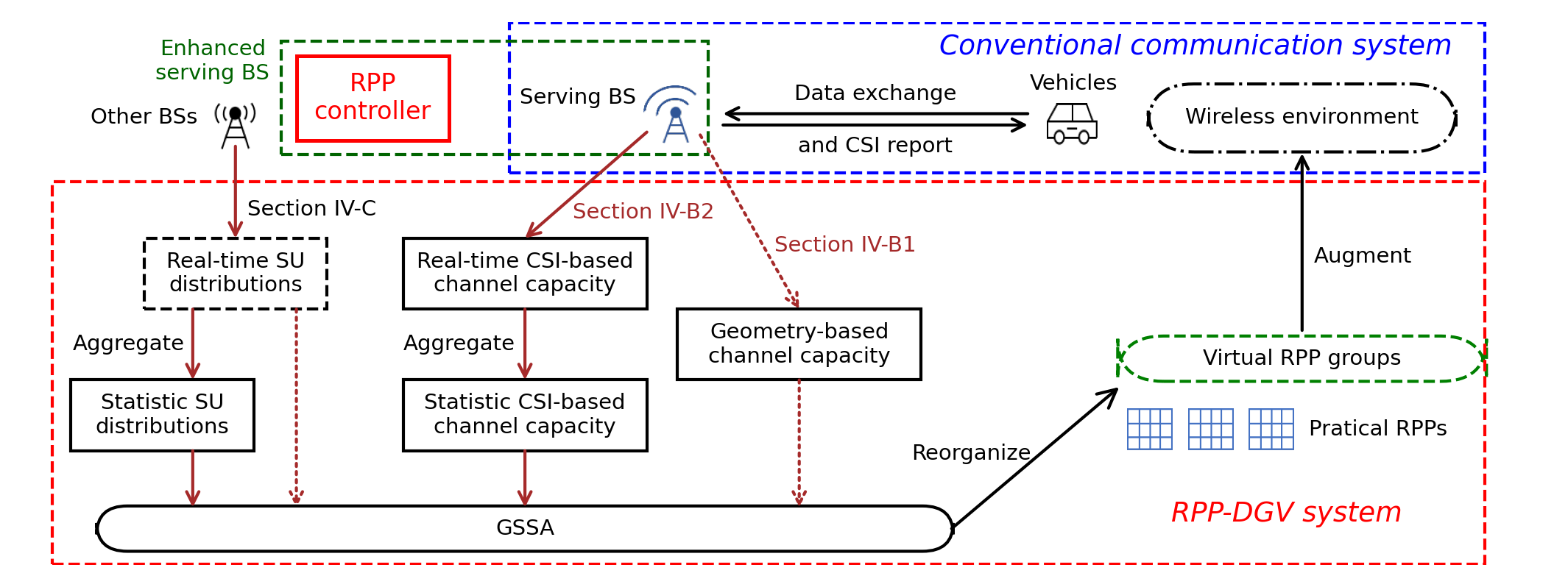}
    \caption{The illustration of RPP dynamic group virtualization (RPP-DGV) system in RETF}
    \label{Framework}
\end{figure*}

\section{Dynamic Deployment of Reflecting Panels}
\label{Dynamic_Deployment}
\subsection{State-Switching Mechanism}
Intuitively, an infinitely long RP can maximize the total throughput of a TV across the entire serving area. However, this will introduce unexpected interference to SUs. Rather than suppressing interference power through complex system designs and phase computations, avoiding scheduling conflicts between the TV and SUs proves more effective. For users served by the same BS $\bm{G}_0$, the serving BS can optimize its scheduling to prioritize the high-throughput transmission of the TV. However, for SUs connected to other BSs, scheduling conflicts become unavoidable, causing severe degradation in SU transmission performance.

To address this problem, we deploy discrete RPs along the road instead of a continuous panel for practical consideration, and adopt a State-Switching Mechanism (SSM). As a TV traverses its route, the RPs can be activated upon the TV's entrance into the corresponding RA and deactivated upon its exit from the RA. Whether stationary or mobile, each SU may encounter interference from reflected signals only during one continuous time interval while co-located with the TV in the same RAs. Given that an SU and the TV can only co-locate in at most two adjacent RAs\footnote{Stationary SU may co-locate in one DRA or two IDRAs. Mobile SU with a different direction from the TV may co-locate in one DRA or IDRA. Mobile SU with the same direction as the TV may co-locate in several discrete DRAs. It can be considered as several independent SUs, and each one may co-locate in one DRA.}, each SU receives the additional reflected interference for a short duration\footnote{Users who co-locate with the TV for a long duration shall be served by the same serving BS and cannot be defined as SUs.}. Thereby, the interference impact on the entire throughput of the SU is mitigated.

The activation and deactivation within an SSM can be implemented in various ways. For instance, an RPP controller can use micro-motors to physically tune the zenith angle $\gamma_i$ of the effective RP $\bm{D}_i$, thereby enabling coverage switching on/off the road. In this paper, we will not specify the implementations but rather use a fixed processing time interval, $T_g$, to represent the time required for an SSM.

With an SSM, the challenging problem of conflict avoidance is reduced to controllable state management of the effective RPs. This capability, combined with an optimal deployment of effective RPs (locations and lengths), can enhance the data transmission of TV without importing too much additional interference to SUs.

\subsection{Virtual Reflecting Panel Patch Groups}
\label{RPP_Groups}
The deployment of RPs is heavily contingent upon the distribution of SUs. The real-time distribution of SUs exhibits significant variability, and even the statistical distributions can fluctuate over time due to the emergence or departure of individual SUs. To effectively manage and minimize interference to SUs in a dynamic manner, we propose the deployment of a series of RPPs along the roadside. This RPP is essentially an independent reflecting panel with a small length and a reduced RA. Then we organize several successive RPPs as one Virtual RPP Group (VRG) to work like one effective RP as illustrated in Fig. \ref{Demo}. These RPPs can be deployed in practical locations, while the VRGs function as a virtual and logical entity managed by the serving BS or RPP controller. All the RPPs in one group are allocated with the same rotation angle and can be activated or deactivated at the same time.

Compared with a regular RP, RPPs can be deployed in challenging locations that may not support larger RPs, resulting in lower deployment and maintenance costs. Additionally, RPPs contribute to increased robustness and flexibility. If some RPPs fail, the remaining ones within the same group can continue to operate, ensuring that the entire RA remains stable for each VRG.

The serving BS can adjust VRGs to meet the requirements of SUs at specific locations in various scenarios. As depicted in Fig. \ref{Demo}, the original VRG comprises six RPPs, with no SUs present initially. However, when new SUs unexpectedly enter the region, they may experience considerable interference while a TV proceeds through it. The BS serving these SUs can inform the BS managing the RPPs to reorganize the VRGs before the TV enters the affected zone. The first three and last two RPPs can be arranged into two independent groups, with the middle RPP, corresponding to the area of new SUs, can be deactivated to minimize the interference. Although the RA miss of an individual RPP may result in a decrease in channel capacity of a TV in that region, this reduction is likely to only marginally affect its throughput. In contrast, reducing additional interference can avoid a severe attenuation in the throughput for the new SUs.



Here, we deploy $N_e$ RPPs along the road, with uniform length of $\lambda_0$ and spacing of $\mu_0$. We assume the distances between these RPPs to the road are the same and constant, expressed as $d_0$. Each RPP can be identified by a unique index $e$ with $e=0,1,2,\dots N_e-1$. For each index $e$, the starting location and the ending location of the corresponding RPP on the X-axis can be indicated with $(\lambda_0+\mu_0)e$ and $(\lambda_0+\mu_0)e +\lambda_0e$, respectively. The VRGs are denoted as
\begin{equation}
    \mathcal{E} = \{\bm{E}_i = \left(e_i^-, e_i^+\right), i=1,2,\ldots, N_E\}
\end{equation}
where $e_i^-$ and $e_i^+$ are the indices of the starting and ending RPPs of VRG $\bm{E}_i$, respectively. $N_E$ is the number of the VRGs. The number of RPPs within $\bm{E}_i$ is denoted as $n_i=e_i^+-e_i^-+1$. Each VRG $\bm{E}_i$ corresponds to an effective RP $\bm{D}_i$ with $P_{i,x}^-=(\lambda_0+\mu_0)e_i^-$ and $P_{i,x}^+=(\lambda_0+\mu_0)e_i^++\lambda_0e_i^+$. The RA of this effective RP is defined as the combination of the RAs of all the RPPs in the group.

For a given location on the overlapping IDRAs of two adjacent RPPs, the sum of two $\omega_{BL}$ can be expressed as
\begin{equation}
    f(\Delta x)=e^{-\chi|\Delta x|^2}+e^{-\chi|\mu_0-\Delta x|^2}
\end{equation}
where $\Delta x$ represents the distance between the given location and the endpoint of the DRA of one RPP. Intuitively, as the spacing $\mu_0$ increases, the total $\omega_{BL}$ decreases. It can be proved that for $\mu_0 < \sqrt{2/\chi}$, the minimum value of $f(\Delta x)$ over the interval $0 \leq \Delta x \leq \mu_0$ is attained at the endpoints and is greater than $1$. With further increase of $\mu_0$, an additional local minimum value appears at the midpoint of IDRA, corresponding to $\Delta x =0.5 \mu_0$. Under the assumption that this local minimum also remains above $1$, the uniform spacing is then constrained by
\begin{equation}
    \mu_0 \leq 2\sqrt\frac{\log2}{\chi}
\end{equation}
With this spacing, when both adjacent RPPs are active, the overlapping IDRAs of these two RPPs can be regarded as DRA with $\omega_{BL}=1$. This indicates that within a group of successive RPPs, the combination of all the RAs of the interior RPPs can be regarded as the DRA of the corresponding effective RP, while the IDRAs of the two endpoint RPPs function as the IDRA of this effective RP.

With the utilization of VRGs, our main objective is to maximize the total throughput of a TV while simultaneously minimizing interference experienced by SUs. Given that the absolute throughput of a TV and each SU can differ significantly in scale, we employ the throughput ratio as a more effective measure than the absolute throughput. The optimization of VRGs is to maximize the joint throughput ratio $\Phi$, expressed as
\begin{equation}
\label{phi}
    \Phi=\Phi_{tar}+\zeta\Phi_{sen}
\end{equation}
where $\zeta$ represents the Interference Level Factor (ILF), indicating the prior weight assigned to SUs. When $\zeta=0$, the influence on SUs is disregarded. As $\zeta$ increases, the significance of interference avoidance becomes more pronounced. $\Phi_{tar}$ and $\Phi_{sen}$ denote the throughput ratios for a TV and SUs, respectively. Here, we define the optimization problem of RPP-DGV as 
\begin{equation}
\label{opt}
\begin{aligned}
    & \left(\bm{\mathcal{P}}\right) \quad \max_{\mathcal{E}} \Phi  \\
    & \space s.t. \quad 0 \leq e_i^{-},e_i^{+} \leq N_e \quad i=1,2\dots N_E \\
    & \space s.t. \quad v_{r}T_g\leq \xi_i \quad i=1,2\dots N_E \\
    & \space s.t. \quad e_i^{+} \leq e_{i+1}^- \quad i=1,2\dots N_E-1
\end{aligned}
\end{equation}

The first condition establishes the fundamental limitations of the value range. The second condition guarantees sufficient time for the VRGs to activate and deactivate. $v_{r}$ denotes the rated speed of the TV, which may be the average speed, the maximum speed, or other designed speeds depending on the demand. The final condition aims to prevent the overlap of VRGs, as this could lead to the reuse of one RPP for multiple RPP groups, which may cause missing coverage. 

One critical work left in the optimization problem $\bm{\mathcal{P}}$ is how to figure out the expressions of the objective function. In the next couple of subsections, we will present the expressions of the throughput ratios $\Phi_{tar}$ and $\Phi_{sen}$, respectively.

\begin{table*}[!t]
    \caption{The influence of different compromised ranges to one sensitive user (SU) $U_j$\tnote{1}}
    \label{Influence_Cases}
    \centering
    \begin{threeparttable}
    \begin{tabularx}{\textwidth}{|>{\hsize=2.4\hsize\centering\arraybackslash}X|*{7}{>{\hsize=0.8\hsize\centering\arraybackslash}X|}}  
        \hline
        Cases & A & \multicolumn{2}{c|}{B} & \multicolumn{3}{c|}{C} & D \\
        \hline
        Sub-cases\tnote{2} & A1 & B2 & B3 & C4 & C5 & C6 & D7 \\
        \hline
        RPP states\tnote{3} & N N N & A N N & N N A & A A N & A N A & N A A & A A A \\
        \hline
        Compromised ranges\tnote{4} & GAP & \multicolumn{2}{c|}{IDRA} & DRA/IDRA & twice IDRA & \multicolumn{2}{c|}{DRA} \\
        \hline
        Endpoint\tnote{5} & \multicolumn{6}{c|}{True} & False \\
        \hline
    \end{tabularx}
    \begin{tablenotes}
    \footnotesize
    \item [1] $U_j$ locates in the DRA or the right IDRA of RPP $e^j$.
    \item [2] Due to the limitation of minimal length, sub-case (N A N) is removed.
    \item [3] The ordinal state of RPP $e^j-1$, $e^j$, and $e^j+1$ where A and N represent the RPP organized in some Virtual RPP Group (VRG) or not, respectively.
    \item [4] To indicate the interference to $U_j$ through reflex channel. GAP means no interference. DRA means the interference is higher with $\omega_{BL}=1$ while IDRA means the interference is smaller with $\omega_{BL} < 1$. Twice IDRA means $U_j$ can be interfered by two VRGs, with a long influence duration $T_j^{(int)}$.
    \item [5] Whether these three RPPs are in the same VRG or there is an endpoint of some VRG.
    \end{tablenotes}
    \end{threeparttable}
\end{table*}

\subsection{Throughput Ratio for TV}
The throughput ratio of a TV pertains to the duration from the moment the TV enters the road until it departs, expressed as
\begin{equation}
\Phi_{tar}=\frac{\int_0^{T}C_{enh}(x_t)dt}{\int_0^{T}C_{org}(x_t)dt}
\end{equation}
where $C_{org}(x_t)$ and $C_{enh}(x_t)$ are the original channel capacities (PAP only) and the enhanced channel capacities (PAP and SAP), respectively.

\subsubsection{Geometry-based channel capacity}
Accurate Channel State Information (CSI) is often unavailable in practice because CSI is unknown before a TV moves onto the road and becomes rapidly outdated due to the fast-varying fading channel. In that case, channel capacity can be calculated with geometrical SINR \cite{MIMO_capacity}, expressed as

\begin{equation}
\begin{aligned}
    & C_{org}(x_t)=WN_r\log_2\left(1+\frac{s_0^{(p)}(x_t)}{\sum_{h=1}^{N_G}s_h^{(p)}(x_t)}\right) \\
    & C_{enh}(x_t)=C_{org}(x_t)+WN_r\log_{2}\left(1+\frac{s_0^{(s)}(x_t)}{\sum_{h=1}^{N_G}s_h^{(s)}(x_t)}\right)
\end{aligned}
\end{equation}
where $W$ is the frequency bandwidth. $s_h^{(p)}(x_t)$ and $s_h^{(s)}(x_t)$ denote the received power from $G_h$ to the PAP and SAP, respectively. 
For $h=0,1,2\dots N_G$, $s_h^{(p)}(x_t) = \overline{s}_h \times\omega_{PL}(\bm{Q}_h,\bm{x}_t) \times g\left(\bm{x}_t,\bm{Q}_h,\bm{B^{(p)}}\right)\times g(\bm{Q}_h,\bm{x}_t,\bm{A_h})$.
For $h=1,2\dots N_G$, $s_h^{(s)}(x_t) = \overline{s}_h \times\omega_{PL}(\bm{Q}_h,\bm{x}_t) \times g(\bm{x}_t,\bm{Q}_h,\bm{B^{(s)}})\times g(\bm{Q}_h,\bm{x}_t,\bm{A_h})$.
For the received power from the serving BS $G_0$ through the reflex channel to SAP, $s_0^{(s)}(x_t) = \overline{s}_0 \times\omega_{PL}^{(ref)}(\bm{Q}_0,\bm{x}_t)\times\omega_{RL}\times\omega_{BL}(x_t) \times g\left(\bm{x_t},\bm{P}_*(\bm{x}_t),\bm{B^{(s)}}\right)\times g\left(\bm{Q}_0,\bm{P}_*(\bm{x}_t),\bm{A_0}\right)$

\subsubsection{CSI-based channel capacity}
\label{CSI-RPP-DGV}

Given that the real-time CSI of a TV is accessible, or if the statistical CSI is acquired from other vehicles, the MIMO channel capacity can be derived as $C=W\sum_{r=1}^{R}\log_2\left(1+\rho_r\sigma_r^2\right)$, where $R$ denotes the total number of layers, which corresponds to the optimal number of orthogonal sub-channels achieved through transmitter precoding. $\rho_r$ represents the receiving SINR for the $r$-th layer, determined by the power allocation at the transmitter. $\sigma_r$ denotes the corresponding singular value that exceeds a predefined threshold. In alignment with 5G cellular systems, we adopt both uniform power distribution across layers and rank adaptation. For the joint channel matrices $H_v, v=1,2\dots N_B$, the singular values are calculated by the average covariance matrix $Cov$, which is expressed as
\begin{equation}
\label{Rank_Adaption}
    Cov=\frac{1}{N_B}\sum_{v=1}^{N_B}\left(\bm{H}_v^{(dir)H}\bm{H}_v^{(dir)}+\omega_*\bm{H}_v^{(ref)H}\bm{H}_v^{(ref)}\right)
\end{equation}
With eigenvalue decomposition of $Cov$, we can get all the effective eigenvalues $\sigma_r^2$. The channel capacity can be calculated as
\begin{equation}
    C_{enh}(x_t)=W\sum_{r=1}^{R}\log_2\left(1+\frac{s_0^{(p)}(x_t)\sigma_r^2(x_t)}{R\sum_{h=1}^{N_G}s_h^{(p)}(x_t)}\right)
\end{equation}
where $\sigma^2_r(x_t)$ denotes the $r$-th largest eigenvalue of $Cov$ at location $x_t$. Similarly, the channel capacity of the original channel $C_{org}(x_t)$ can also be calculated based on $H_v^{(dir)}$ only.


It is evident that the CSI-based channel capacity, which accounts for the combination of two channel matrices, more closely approximates the real communication environment. Consequently, the corresponding CSI-based RPP-DGV can achieve a better performance balance between the TV and SUs. Since the channel capacity is independently computed for each location, the RPP may employ a hybrid approach. It can initialize the RPP-DGV using geometry-based channel capacities and refine the virtualization by updating to CSI-based RPP-DGV for locations where CSI becomes available.

\subsection{Throughput Ratio of Sensitive Users}
As discussed in Section \ref{Dynamic_Deployment}, each SU may be influenced by only one or two VRGs. The indices of these VRGs are based on the location and speed of an SU. We represent these indices with the set $\bm{I}_j$, which contains at most two components. If $\bm{I}_j=\varnothing$, it indicates that $\bm{U}_j$ is located in the gap between two effective RPs and may not experience interference from any effective RP. The total interference duration of $\bm{U}_j$ can be expressed as
\begin{equation}
    T_j^{(int)}=\sum_{i \in \bm{I}_j}\min(\frac{\xi_i}{v_{r}},\frac{\psi_{i,j}}{v_{j}})
\end{equation}
where $\psi_{i,j}$ denotes the distance between $\bm{U}_j$ and the corresponding endpoint of RPP group $\bm{E}_i$ when it is activated. The average throughput ratio of the SUs can be expressed as
\begin{equation}
    \Phi_{sen}=\frac{1}{N_U}\sum_{j=1}^{N_U}\left(1-\frac{T_j^{(int)} \left(C_{nrm}^{(j)}(x_j)-C_{int}^{(j)}(x_j) \right)}{\int_0^{T}C^{(j)}_{nrm}(x_j+v_j(t-t_j))dt}\right)
\end{equation}
where $t_j=x_j/v_r$ indicates the time when the TV and $\bm{U}_j$ encounter. $C_{nrm}^{(j)}(x_t)$ and $C_{int}^{(j)}(x_t)$ represent the normal channel capacity and the interfered channel capacity of $\bm{U}_j$ at location $x_t$, respectively. Given that the location change of $\bm{U}_j$ in duration $T_j^{(int)}$ is much smaller than the distance to its serving transmitter, we use the capacity difference at location $x_t=x_j$ to represent the interference impact. 

Given that these SUs are not associated with the serving BS, the RPP controller can only use the geometrical information updated by their own serving transmitters to estimate their channel capacities, expressed as
\begin{equation}
\begin{aligned}
    & C_{nrm}^{(j)}(x_t) = W\log_2\left(1+\frac{s^{(j)}_{h_j}(x_t)}{s^{(j)}_0(x_t)+n^{(j)}(x_t)}\right) \\
    & C_{int}^{(j)}(x_t) = W\log_2\left(1+\frac{s^{(j)}_{h_j}(x_t)}{s^{(j)}_0(x_t)+r^{(j)}_0(x_t)+n^{(j)}(x_t)}\right)
\end{aligned}
\end{equation}
where $n^{(j)}(x_t)=\sum_{h=1,h\neq h_j}^{N_G}s^{(j)}_{h}(x_t)$ represents the total interference power from transmitters except $\bm{G}_0$. 
$s^{(j)}_{h}(x_t)=\overline{s}_h\times \omega_{PL}(\bm{Q}_h, \bm{x}^{(j)}_t)\times g(\bm{Q}_h, \bm{x}^{(j)}_t,\bm{A}_h)$ denotes the received power of $\bm{U}_j$ from $\bm{G}_{h}$, where $\bm{x}^{(j)}_t=(x_t,y_j,0)$ indicates the 3-D coordinate of $\bm{U}_j$ at time $t$. $r^{(j)}_{0}(x_t)=\overline{s}_0 \times \omega_{PL}^{(ref)}(\bm{Q}_0,\bm{x}^{(j)}_t)\times \omega_{RL}\times \omega_{BL}(x_t) \times g\left(\bm{Q}_0, \bm{P}_*(\bm{x}_t),\bm{A}_0\right)$ represents its received power from $\bm{G}_0$ via reflex channel.

\subsection{Greedy Stepwise Search Algorithm (GSSA)}
The critical step in putting RPP-DGV into practice is to solve the optimization problem $\bm{\mathcal{P}}$. While various methods can be utilized for this optimization, they typically require a considerable amount of time for iterations. To facilitate the swift reorganization of RPPs in response to dynamic changes in the distribution of SUs or fluctuations in the wireless channel environment, we propose a greedy method, termed as GSSA, which evaluates each SU independently, enabling a more efficient and prompt resolution of the optimization challenge.

Each RPP exists in one of two states: either it is organized in a specific VRG or it is not. Due to the large number of RPPs, it is impractical to explore all possible states of $N_e$ RPPs to identify the optimal solution through exhaustive search. However, the number of SUs is usually considerably smaller. For each SU $U_j$, its location is in the RA of an RPP with index $e^j$. $U_j$ can only be influenced by three neighboring RPPs with indexes $e^j-1$, $e^j$, and $e^j+1$. Consequently, there are at most eight distinct state combinations for these three RPPs, as outlined in Table \ref{Influence_Cases}.

\begin{algorithm}[H]
    \caption{Greedy Stepwise Search Algorithm (GSSA)}
    \label{GSSA}
    \begin{algorithmic}
       \STATE{\bf{Initialization}}
       \STATE{\hspace{0.4cm} Calculate the minimal number of RPPs in one group if needed. Otherwise, $n_0=0$.}
       \STATE{\hspace{1.0cm} $n_0=V_{r}T_g/(\lambda_0+\mu_0)$}.
       \STATE{\hspace{0.4cm} Sort SUs in $\mathcal{U}$ by increasing order of the distance to $\bm{G}_0$.}
       \STATE{\hspace{0.4cm} Generate the initial VRGs.}
       \STATE{\hspace{1.0cm} $\mathcal{E}^{(0)}=\{\bm{E}_1=(0,N_e-1)\}$.}
       \STATE{\hspace{0.4cm} Calculate the initial joint throughput ratio $\Phi^{(0)}$.}
       \STATE{\textbf{For} each $\bm{U}_j$ in $\mathcal{U}$, \textbf{do}}
       \STATE{\hspace{0.4cm} Find the three corresponding RPPs $e^j-1$, $e^j$, and $e^j+1$.}
       \STATE{\hspace{0.4cm} \textbf{For} each sub-cases in Table \ref{Influence_Cases}, \textbf{do}}
       \STATE{\hspace{1.0cm} Reorganize the VRGs $\mathcal{E}$.}
       \STATE{\hspace{1.0cm} Calculate the joint throughput ratio $\Phi$.}
       \STATE{\hspace{1.0cm} \textbf{If} $\Phi<\Phi^{(j-1)}$ or $\exists \bm{E}_i$ with $n_i<n_0$}
       \STATE{\hspace{1.5cm} Remove this sub-case.}
       \STATE{\hspace{1.0cm} \textbf{End If}}
       \STATE{\hspace{0.4cm} \textbf{End For}}
       \STATE{\hspace{0.4cm} \textbf{If} there remain sub-cases, select the optimal one}
       \STATE{\hspace{1.5cm} $\mathcal{E}^{(j)} \xleftarrow{} \mathcal{E}$ and $\Phi^{(j)} \xleftarrow{} \Phi_{\max}$}
       \STATE{\hspace{0.4cm} \textbf{Else}}
       \STATE{\hspace{1.5cm} $\mathcal{E}^{(j)}$ and $\Phi^{(j)}$ keep unchanged.}
       \STATE{\hspace{0.4cm} \textbf{End If}}
       \STATE{\textbf{End For}}
    \end{algorithmic}
\end{algorithm}

By exploring the seven sub-cases associated with the three designated RPPs, we can identify the optimal VRGs for $U_j$. We begin with a single VRG encompassing all RPPs and subsequently analyze all SUs to determine the optimal number and combination of VRGs, as depicted in Algorithm \ref{GSSA}. When SUs are distributed sparsely, this algorithm can achieve optimal results. In cases where there are a large number of SUs and multiple users may be affected by a single VRG, the algorithm is still capable of producing a sub-optimal outcome. It can work as a RPP-DGV algorithm or serve as a starting point for other iteration-based optimization methods.

To effectively address the dynamic distribution of SUs, the GSSA can be utilized based on the current state. When a SU is no longer present, the associated three RPPs can be reassigned to the nearest VRG. Conversely, when new SUs are introduced, the current organization of VRGs can be treated as an initial starting point with a joint throughput ratio $\Phi^{(0)}$. GSSA can then be applied directly to these new users. This strategy enables the RPP controller to swiftly reorganize the VRGs in response to any changes in the distribution of SUs.

\subsection{Reflection Enhanced Transmission Frame (RETF)}
Now, we are ready to present our framework for the reflection-enhanced data transmission, named as RETF, as depicted in Fig. \ref{Framework}. RETF consist of two distinct systems, the RPP-DGV system and a conventional communication system.

Within the RPP-DGV system, the RPP controller dynamically switches between geometry-based and CSI-based RPP-DGV based on whether CSI is available, as discussed in Section \ref{CSI-RPP-DGV}. Since CSI-based channel capacities are time-sensitive, the controller will disregard any obsolete records. If insufficient current information is available, similar to the initial deployment phase, the controller reverts to geometry-based RPP-DGV.
The RPP-DGV system implements a dynamic strategy where the controller selects between geometry-based and CSI-based RPP-DGV based on CSI availability as discussed in Section \ref{CSI-RPP-DGV}. Owing to the time-sensitive nature of CSI-based channel capacities, any obsolete records are discarded. In cases where current CSI is insufficient, the system reverts to the geometry-based RPP-DGV, similar to its initial deployment state.


A defining feature of the RPP-DGV system is its operational independence from conventional communication systems. It augments the wireless environment without controlling the data exchange procedures, thereby enabling rapid adoption within any existing system. The RPP controller can function as a standalone entity, managing RPPs across a designated area for all communication systems. Alternatively, it can be integrated into a serving BS, enhancing it to organize RPPs for the performance improvement of a specific system, as illustrated in Fig. \ref{Framework}.

\section{Rotational Collaboration of virtual RPP Groups}
In general, the received power over a reflex channel is lower than that of the direct channel due to increased path loss caused by the longer transmission distance and RL. The received power may still be inadequate even when a TV is situated within DRAs, where $\omega_{BL}=1$. In the procedure of rank adaptation, the eigenvalues are calculated based on \eqref{Rank_Adaption} where $ \omega_* $ corresponds to the power gap between PAP and SAP. 

Given that $N_r \ll N_t$, the covariance matrices of both the direct channel and the reflex channel can be considered low-rank matrices. When the reflex channel and the direct channel are orthogonal, the eigenvalues of the joint channel matrix can be interpreted as a combination of the eigenvalues from these two channels. The small value of $\omega_*$ can exacerbate the eigenvalue differences among layers. Since transmit power is uniformly distributed across each layer at the transmitter, the maximum SINR is attained when all eigenvalues are equal. Greater disparities among the layers can result in power inefficiencies and reduced effective SINR. Consequently, even if the overall channel is high-rank, the transmitter typically opts for a smaller number of layers during the rank adaptation process.

To enhance the received power of the reflex channel, we propose a signal focusing method by utilizing physical rotation to adjust the orientations of VRGs in a specific team, termed as RCT. This technique enables multiple VRGs to concurrently reflect signals to the same TV, just like the magnifying glass to focus weak sunlights to the burning point, thereby increasing the overall received power.

\begin{figure}
    \centering
    \includegraphics[width=3.4in]{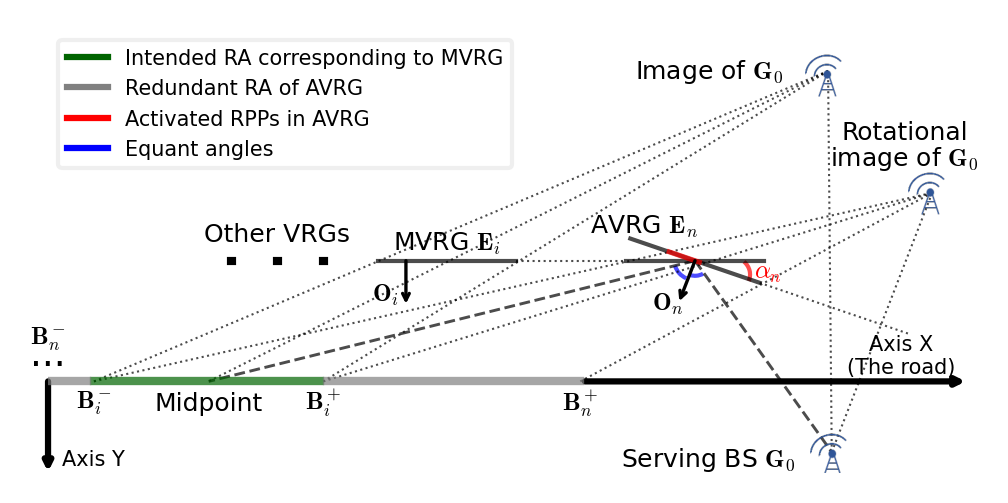}
    \caption{The midpoint calibration between two VRGs in the rotational collaboration team (RCT)}
    \label{Rotation}
\end{figure}

\subsection{Rotational Collaborative Team (RCT)}
The RCT consists of multiple VRGs. When a TV is positioned in a specific RA, $\bm{F}_i$, the corresponding VRG without rotation, $\bm{E}_i$, is recognized as the Major Vritual RPP Group (MVRG). The remaining groups are classified as Associated Virtual RPP Groups (AVRG). Within each RCT, there is one MVRG and several MVRGs, determined by the RA in which the TV is situated.

To ascertain the rotation angles of the AVRGs corresponding to a designated MVRG $\bm{E}_i$, we utilize the RA midpoint calibration technique, as demonstrated in Fig. \ref{Rotation}. For simplicity, the RA of each VRG is approximated by the DRA of all its constituent RPPs. The rotation angle for each AVRG is then obtained by aligning its RA midpoint with that of the MVRG, expressed as
\begin{equation}
    \alpha_n=\arctan{\frac{(\sigma-1)h + h_0}{(\sigma-1)d_0 - q_0}}
\end{equation}
where $h_0$ and $h$ denote the distances from the serving BS $\bm{G}_0$ and the given AVRG $\bm{E}_n$ to the MVRG $\bm{E}_i$ along the x-axis, respectively. $q_0$ indicates the distance from $\bm{G}_0$ to the road. $\sigma$ represents the distance factor, expressed as 
\begin{equation}
    \sigma=\frac{\sqrt{(h-h_0)^2+(d_0+q_0)^2}}{\sqrt{h^2+d_0^2}}
\end{equation}
RPP rotation leads to misalignment relative to the road. Consequently, even if the midpoints of RAs coincide, the effective RA projected onto the road for each AVRG differs from its non-rotated counterpart. This results in redundant regions $[B_n^-, B_i^-]$ and $[B_i^+, B_n^+]$, as illustrated in Fig. \ref{Rotation}, which provide no intended benefit and risk causing interference to other SUs. To address this, the specific RPPs responsible for these redundant regions must be selectively deactivated in each AVRG.


For a given RCT and the MVRG $\bm{E}_i$, the first index of the active RPP in AVRG $\bm{E}_n$ corresponds to the largest index within the range $[e_n^-,e_n^+]$ that fulfills the condition:
\begin{equation}
    e_*^- + \frac{d_0}{\tan{(2\alpha_n-\beta_e)}} \leq \frac{(q_0+2d_0)(\lambda_0+\mu_0)e_i^--q_0d_0}{(q_0+d_0)(\lambda_0+\mu_0)}
\end{equation}
where $\beta_e \in [0, \pi]$ represents the angle related to $\bm{G}_0$, expressed as
\begin{equation}
    \beta_e=\arctan{\frac{q_0+d_0}{Q_{0,x}- (\lambda_0+\mu_0)e_*^-}}
\end{equation}
where $e_*^-$ corresponds to the left endpoint of the effective RP, and the right endpoint can be similarly represented based on $(\lambda_0+\mu_0)e_i^++\lambda_0$. This method enables the RPP controller to selectively rotate specific RPPs within a single AVRG, thereby covering the entire RA without introducing interference in adjacent areas.



\begin{figure}
    \centering
    \includegraphics[width=3.4in]{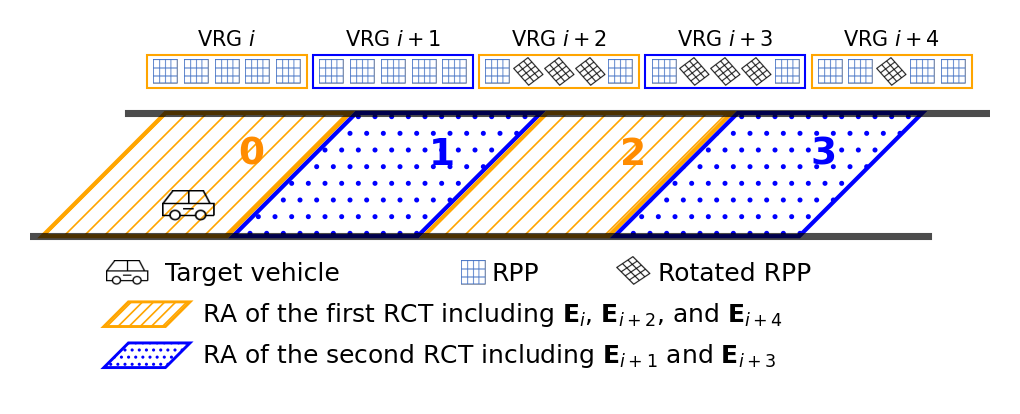}
    \caption{The alternating neighbor selection (ANS) of RCTs}
    \label{Collaboration}
\end{figure}

\subsection{Alternating Neighbor Selection}
In a specific RCT, the movement of a TV can trigger a switch of MVRG, resulting in variations in the rotation parameters of one particular AVRG, including $\alpha_i$ and the valid RPP indexes. Typically, all rotation parameters are calculated and stored once the RPP-DGV is complete. However, when the TV is about to move into some specific RA, BS requires sufficient time to finish the rotation of the RPPs within each AVRG of the RCT. Consequently, adjacent RPP groups cannot operate simultaneously within a single RCT, meaning that one RCT cannot effectively serve adjacent RAs.

To ensure full road coverage, we introduce a method for selecting RCTs, named Alternating Neighbor Selection (ANS). This method requires at least two RCTs, as illustrated in Fig. \ref{Collaboration}. We assign $\bm{E}_{i}$, $\bm{E}_{i+2}$, and $\bm{E}_{i+4}$, to one RCT, while $\bm{E}_{i+1}$ and $\bm{E}_{i+3}$ are allocated to another RCT. When the TV is in region $0$, $\bm{E}_{i}$ serves as the MVRG, with $\bm{E}_{i+2}$ and $\bm{E}_{i+4}$ functioning as AVRGs. Meanwhile, $\bm{E}_{i+1}$ and $\bm{E}_{i+3}$ are preparing for the subsequent region $1$. The RPPs in $\bm{E}_{i+1}$ are transitioned to a non-rotation state, while some RPPs in $\bm{E}_{i+3}$ are adjusted to orient towards region $1$. If $\bm{E}_{i+1}$ were to act as an AVRG alongside $\bm{E}_{i}$ for region $0$, there would not be sufficient time for $\bm{E}_{i+1}$ to properly adjust its orientation as the TV moves into region $1$. Similarly, when the TV is in region $1$, $\bm{E}_{i+2}$ is switched to a non-rotation state to cover Region $2$, while $\bm{E}_{i}$ and $\bm{E}_{i+4}$ are rotated to take on associated roles in covering Region $2$. Thus, the TV will be enhanced by these two RCTs alternately throughout its entire path.

\section{Performance Evaluation}
To align our evaluation with the real 5G cellular systems, we develop a System-Level Simulation (SLS) platform in accordance with the 3GPP RAN1 guidelines specified in TR 38.901 \cite{3GPP901} and TR 36.889 \cite{3GPP889}. We ensure its accuracy through channel model calibration and MIMO calibration based on configurations and results in \cite{3GPP901,3GPPcm}. Integrated with the reflecting channel model from Section \ref{Model}, this platform enables the implementation of RETF to assess the performance of both TVs and SUs.


In our simulation, the serving BS is equipped with 32 antennas and positioned at the midpoint of the road along the X-axis. The PAP and SAP of a TV are placed opposite each other with 180 degrees separation. CSI feedback is implemented through the enhanced Type II codebook standardized in 3GPP TS 38.214 \cite{3GPP214}. The serving BS directly utilizes the reported CSI to compute the precoding matrix and allocate all frequency bands to TV. Dynamic rank adaptation and Minimum Mean Square Error with Interference Rejection Combining (MMSE-IRC) receiver are adopted for both TVs and SUs. For generality, throughput is quantified by spectral efficiency (SE), measured in bit/s/Hz.


\begin{figure*}
    \centering
    \subcaptionbox{SE of the TV and SUs on different locations along the road\label{eval_tv}}[0.74\textwidth]{\includegraphics[width=\linewidth]{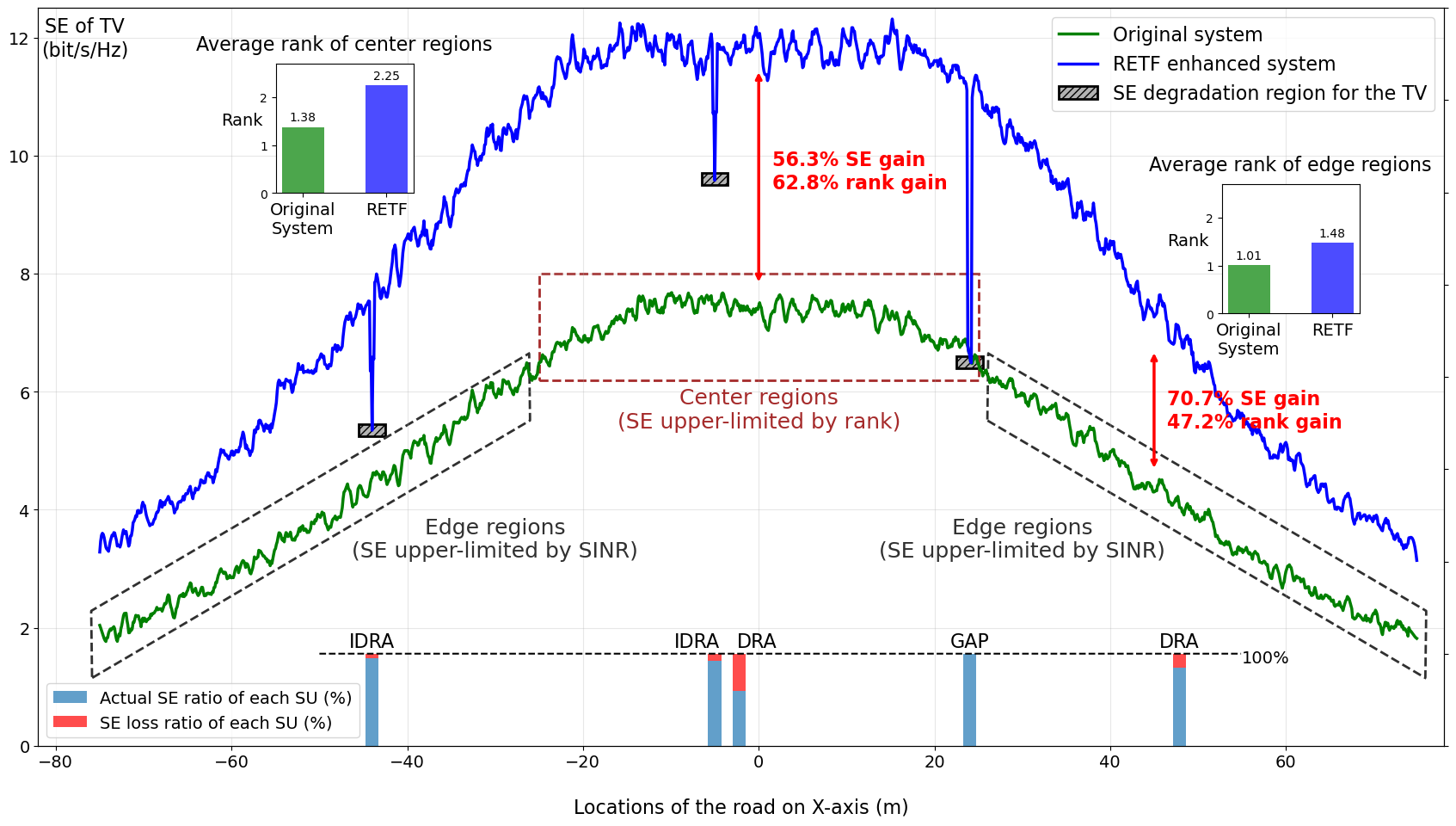}}
    \subcaptionbox{Influence of total SU numbers\label{eval_su}}[0.25\textwidth]{\includegraphics[width=\linewidth]{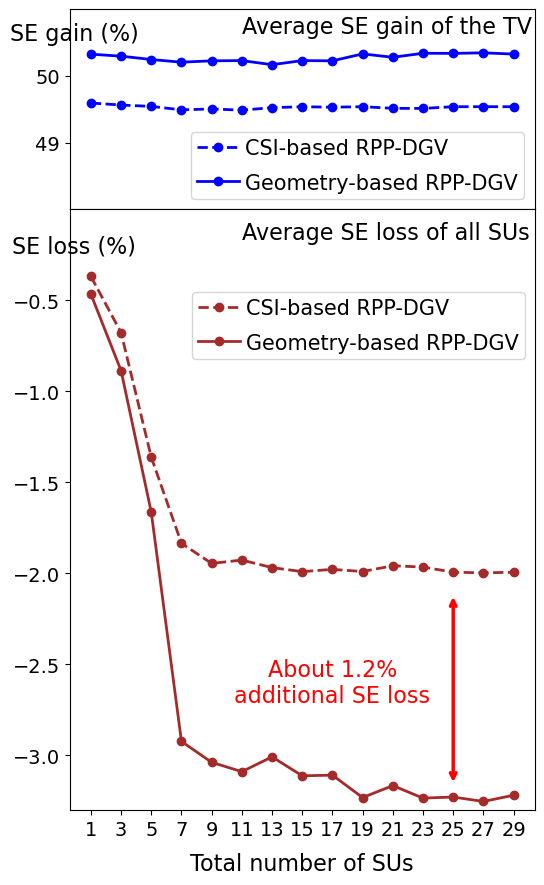}}
    \caption{Performance of the target vehicle (TV) and sensitive users (SU)}
    \label{Evaluation1}
\end{figure*}

\begin{figure*}
    \centering
    \subcaptionbox{Center regions\label{eval_center}}[0.49\textwidth]{\includegraphics[width=\linewidth]{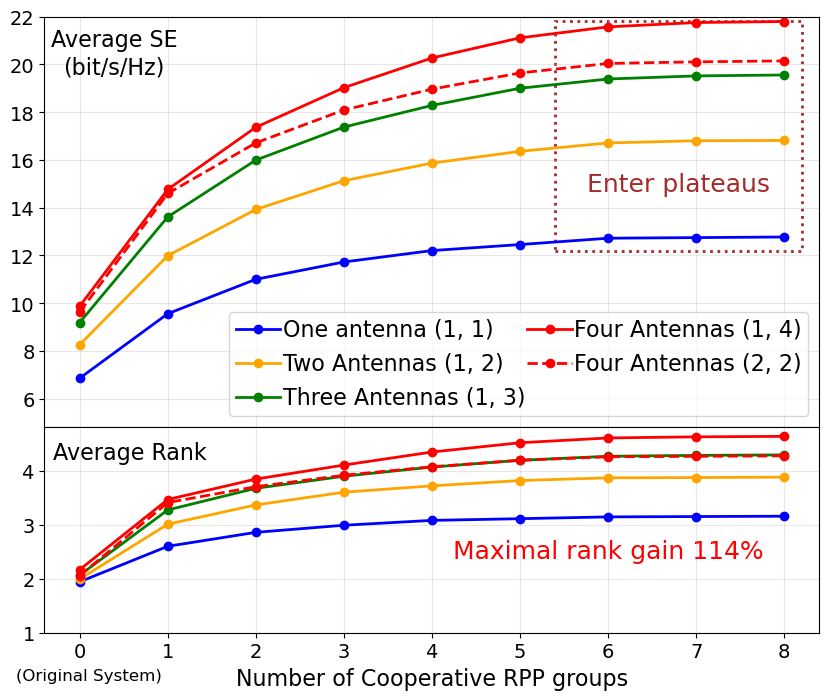}}
    \subcaptionbox{Edge regions\label{eval_edge}}[0.49\textwidth]{\includegraphics[width=\linewidth]{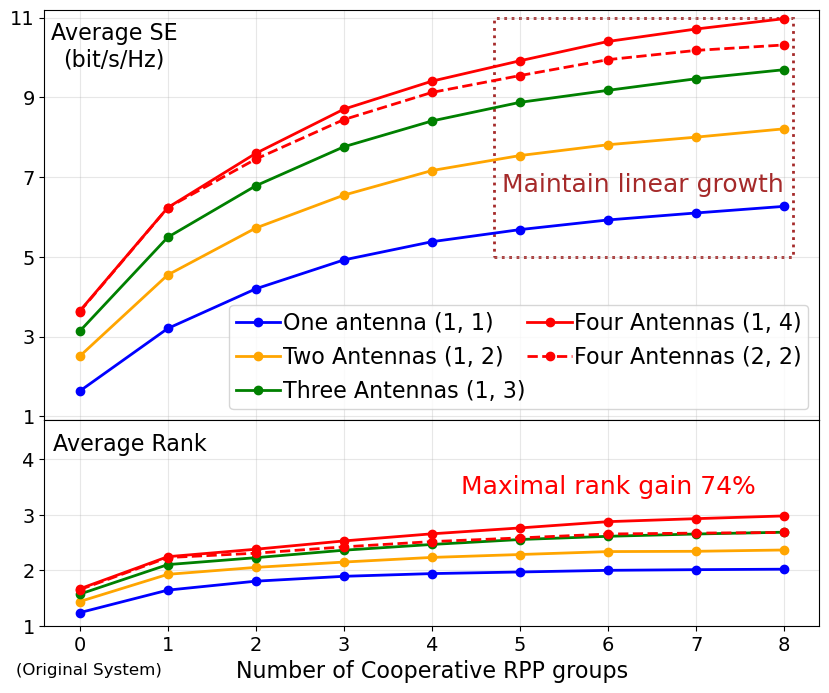}}
    \caption{Average SE and rank curves with different VRG numbers in each RCT\protect\footnotemark}
    \label{Evaluation2}
\end{figure*}

The performance evaluation of a TV equipped with two dual-polarized antennas on its PAP and SAP is illustrated in Fig. \ref{eval_tv}. In center regions, where the relatively high SINR drives the maximum MCS, the throughput ceiling is predominantly constrained by the achievable channel rank. Compared to the single direct path in conventional systems, the RETF scheme introduces additional reflected paths, enhancing channel diversity. This diversity elevates the average channel rank from 1.38 to 2.25, facilitating the transmission of a greater number of parallel data streams over identical time–frequency resources and thereby significantly boosting the achievable SE. Considering the maximal sub-6 GHz bandwidth, the peak transmission rate reaches approximately 1.2 Gbps.

In contrast, edge regions suffer from low SINR due to weak received signals and strong interference, making throughput ceiling constrained by SINR rather than rank. Under these conditions, scheduling additional layers diminishes the transmit power per layer, making interference suppression more critical than rank enhancement. In RETF, the presence of reflected paths from approximately opposite directions enables a TV to utilize additional antennas on SAP, thereby enhancing the interference cancellation capability of the MMSE-IRC receiver. Consequently, even if the channel rank in RETF remains below 1.5, the improved SINR can yield an SE gain exceeding 70\%.

Notably, as shown in Fig. \ref{eval_tv}, within the compromised ranges (DRA, IDRA, and GAP), the impacts on the SE of both a TV and SUs are temporary. These SE degradations are slight over the entire observation period, due to the SSM. Fig. \ref{eval_su} further shows that, with an increasing number of SUs, the average SE loss for SUs is contained below 3\%, while the TV’s SE gain remains stable. Compared to the CSI-based RPP-DGV, the Geometry-based RPP-DGV incurs an additional 1.2\% SE loss to SUs. This is attributed to the lack of channel information required for accurate interference estimation from reflected paths. Hence, the CSI-based RPP-DGV is preferred when sufficient CSI is available. Otherwise, during the procedures of Geometry-based RPP-DGV, an empirical power factor can be introduced into the calculation of $\Phi_{sen}$ to compensate for the underestimated interference.

Next, we investigate the performance of RETF under different antenna structures and RCT sizes (VRG numbers in RCT). As shown in Fig. \ref{Evaluation2}, the average SE increases significantly with the RCT size for both center and edge regions. In center regions, however, both SE and rank improvements saturate when RCT size exceeds six, as additional AVRGs are too distant from the MVRG to contribute effectively to channel diversity. In contrast, within edge regions, SE continues to increase nearly linearly with RCT size despite limited rank improvement, since even small power gains are crucial for SINR improvement. As quantified in Table \ref{comparison}, the SE gain from an 8-VRG RCT over a 1-VRG RCT is substantially larger in edge regions (118.3\%) than in center regions (70.8\%). Consequently, a practical strategy is to use smaller RCTs (fewer than six groups in our simulation) in center regions to reduce interference to SUs, whereas larger RCTs can be employed in edge regions to maximize SE, thereby balancing performance and interference across the entire road.

\begin{table}[htbp]
    \centering
    \caption{The comparison of 4-antennas performance between the maximal and minimal RCT size}
    \label{comparison}
    \begin{tblr}{
        width = \columnwidth,  
        colspec = {c c c c c}, 
        cell{1}{1} = {r=2}{c}, 
        cell{1}{2} = {c=2}{c}, 
        cell{1}{4} = {c=2}{c}, 
        hlines, vlines,        
        rows = {m},            
        row{1} = {font=\bfseries}, 
    }
    \centering RCT size & \SetCell[c=2]{c} Center regions &  & \SetCell[c=2]{c} Edge regions &  \\
                  & Rank & SE & Rank & SE \\
    1 VRG   & 59.9\% & 49.5\% & 34.8\% & 76.0\% \\
    8 VRGs  & 113.4\% & 120.3\% & 71.2\% & 194.3\% \\
    Delta  & 53.5\% & \textbf{70.8\%} & 36.4\% & \textbf{118.3\%} \\
    \end{tblr}
\end{table}

\footnotetext{The structures of dual-polarized antenna in the legends apply identically to both PAP and SAP as discussed in Section. \ref{LOS_Angle}.}

According to Fig. \ref{Evaluation2}, SE gain also increases with the number of antennas due to the enhanced channel diversity and interference cancellation capabilities provided by a larger antenna array. A comparison of the solid and dashed red curves indicates that a horizontal antenna configuration yields greater performance gains than a vertical one. This result aligns with our simulation geometry, where all RPPs are distributed and rotated primarily within the horizontal plane. In practice, deploying RPPs in both horizontal and vertical dimensions would allow the serving BS to dynamically select the most suitable RPPs based on a TV's specific antenna configuration during the RPP-DGV procedure. This strategy could further improve SE with minimal additional RPP deployment overhead.



\section{Conclusion}
In this paper, we have investigated the throughput ceiling of individual vehicle in conventional communication systems under low-frequency, high-mobility, single-link scenarios, based on the analysis of the link sensitivity and traffic burstiness requirements of emerging intelligent vehicle applications. To break this throughput ceiling, we have proposed a novel solution employing dedicated specular reflecting surfaces to provide various radio paths. These additional radio paths come from diverse directions towards a target vehicle, thereby activating additional antennas and enhancing spatial diversity. Analogous to the way a concave mirror reflects and focuses light onto a single point, this approach coherently integrates signals to improve the rank of the joint channel matrix, ultimately enabling the serving BS to allocate more layers to the target vehicle. In this study, we have deployed a series of reflecting panel patches along the road and organized contiguous patches into a virtual group, allowing them to function as a single effective reflecting panel. By utilizing dynamic virtualization and state-switching mechanism, we can maximize the transmission rate of a target vehicle while simultaneously minimizing interference experienced by other existing sensitive users. We have proposed both CSI-based and geometry-based virtualization techniques to tackle scenarios involving known and unknown CSI, respectively. Moreover, we have proposed a novel reflection-enhanced transmission framework that is applicable to any existing communication system. In the framework, a fast optimization algorithm is proposed to support the real-time dynamic reorganization of reflecting panel patches. We have also presented the rotational collaborative team and alternating neighbor selection method to further enhance the reflected signal convergence. Finally, we have assessed the performance of our framework using the system-level simulations provided by 3GPP, and our findings indicate that a substantial improvement in throughput ceilings can be realized as anticipated.


\bibliographystyle{IEEEtran}
\bibliography{ref}

\end{document}